# Effect of Crystallographic Orientation on the Potential Barrier and Conductivity of Bessel Written Graphitic Electrodes in Diamond

Akhil Kuriakose[a,b], Andrea Chiappini[c], Pietro Aprà[d], and Ottavia Jedrkiewicz[a]

a. Istituto di Fotonica e Nanotecnologie (IFN)-CNR, Udr di Como, Via Valleggio 11, 22100 Como, Italy; akuriakose@studenti.uninsubria.it, ottavia.jedrkiewicz@ifn.cnr.it
b. Dipartimento di Scienza e Alta Tecnologia, Università dell'Insubria, Via Valleggio 11, 22100 Como, Italy;
c. Istituto di Fotonica e Nanotecnologie (IFN)-CNR, CSMFO and FBK Photonics Unit, Trento, Italy; andrea.chiappini@ifn.cnr.it
d. Department of Physics and "NIS" Inter-departmental Centre, University of Torino National Institute of Nuclear Physics, sect. Torino, Via Pietro Giuria 1, 10125, Torino, Italy;

Correspondence: ottavia.jedrkiewicz@ifn.cnr.it

# Abstract

Ultrafast laser micromachining can be used to promote diamond graphitisation, enabling the creation of electrically conductive wires embedded in the diamond matrix. In this context, the presence of a potential barrier in the conductivity of transverse graphitic wires fabricated by pulsed Bessel beams without sample translation across 500 μm thick monocrystalline CVD diamond has been studied. In particular, the role of the crystallographic orientation has been analysed. The morphology and the conductivity of the obtained electrodes have been studied using optical microscopy and current-voltage measurements, while the structural changes have been investigated by means of micro-Raman spectroscopy. By using different laser writing parameters, we have explored the features of different electrodes in a (100) and a (110) oriented diamond crystal respectively. We show that in addition to the use of specific pulse energies and durations (in the fs and ps regimes), the crystallographic orientation of the sample plays an important role in reducing or eliminating the potential barrier height of the IV electrical characterization curves. In a (110) oriented sample, it is possible to eradicate the potential barrier completely even for graphitic wires fabricated at low pulse energy and in the fs pulse duration regime, in contrast to the (100) oriented-crystal case where the barrier is generally observed.

The effect of thermal annealing of the diamond samples on the resistivity of the fabricated micro-electrodes has also been investigated. In (110) oriented diamond, resistivities lower than 0.015 Ω cm have been obtained.

# 1. Introduction

Diamond is an excellent material platform thanks to its unique properties such as high thermal conductivity, good biocompatibility, high radiation stability, remarkable hardness and high chemical resistivity [1, 2]. Moreover, the presence of NV centres in diamond has attracted a lot of attention due to their potential applications in quantum communication and sensing at nanoscale regime [3]. Another prominent feature of diamond is that it can be converted to graphite-like carbon thus creating conductive channels within a sample that is insulating in nature [4]. Such electrodes, which are used as electric field generators, find numerous applications in photonic chips [5], radiation detectors [6], microfluidic sensing systems [3] and biosensing [7]. There are many techniques for conductive electrode fabrication in diamond such as thermal annealing, ion beam lithography and laser writing. Annealing the diamond sample under vacuum around 1700º C can lead to the formation of a graphitic layer on the surface resulting in the generation of a conductive region independent of surrounding gas pressure [8]. Ion beam lithography operates by irradiating diamond with ion beams at keV or MeV energy levels, promoting the creation of lattice defects. These defects enable the transformation of the damaged region into graphite if a critical threshold (graphitization threshold) [9-10] is exceeded.

Fabrication of microstructures in different transparent materials using ultrafast lasers has emerged as an excellent tool for the precise modification of the same. Thanks to the multiphoton absorption and avalanche ionization, the non-linear absorption of the ultrashort pulses helps in avoiding collateral damages due to thermal effects thus resulting in a modification of the material within a confined volume [11]. This spatial confinement, together with laser-beam scanning or sample translation, makes it possible to fabricate different microstructures in three dimensions within the bulk of the material. In diamond, a conventional Gaussian beam can be used for the optical breakdown of the material due to the local transformation of diamond into graphite at the focused beam waist. As a result, a conducting graphitized region is formed in the insulator and a uniform translation of the laser focus throughout the crystal induces the formation of a wire-like graphitized region with a length of our choice [12]. The optical breakdown generates one or more initial graphitic seeds inside diamond and the modified region continues as an extension toward the laser beam as a graphitization wave thanks to the good absorption of radiation by graphite. Taking advantage of this graphitization wave, it is possible to create conductive microstructures within the diamond bulk in any orientation [13].

Recently, finite energy Bessel beams [14] (i.e., Bessel–Gauss beams) which are characterised by a cylindrical symmetry and a transverse profile featured by an intense central spot surrounded by weaker concentric rings resulting in an elongated non-diffracting zone have been utilised for in-bulk modification of transparent materials. Thanks to this elongated focal length (also known as Bessel Zone) where the central beam core is resistant to diffraction spreading, sample translation along the beam propagation direction for in-depth microstructuring is not required in contrast to the standard laser processing techniques using focused Gaussian pulses [15-16]. Due to their self-reconstruction property and their elongated focal zone, finite energy Bessel beams are ideal not only for internal microstructuring but also for high-impact technological applications such as high-speed cutting, welding or drilling of transparent materials [17-20].

Fabrication of graphitic microwires in diamond bulk using pulsed Bessel beams was initially reported by Canfield et al. [21], and a detailed study of the resulting micro-electrodes as a function of the laser writing parameters was recently presented by our group in [22], where resistivity values on the order of 0.04 $\Omega$ cm were obtained thanks to an optimization of the Bessel pulse energy and duration. In there, the presence of a potential barrier in the electrical measurements related to the graphitic wires generated in the femtosecond regime was shown but not addressed in a detailed manner. The potential barrier refers to the voltage gap only after which the electrode shows a perfectly ohmic behaviour. Its presence affects the quality of the electrode and to have good conductivity values, it is of utmost importance to avoid such barriers. The presence of a potential barrier in laser written graphitic wires fabricated with Gaussian beams has been addressed in [23] for a (100) oriented diamond, as a function of different laser writing parameters. In contrast, here, in this work, where the graphitic electrodes are fabricated without sample translation nor beam aberration correction techniques, using the advantage of an elongated focus of the Bessel beam injected orthogonally to the top surface of the samples, we show that in addition to the use of specific pulse energies and durations (both in the fs and ps regimes), the crystallographic orientation of the sample (i.e. having a (100) or (110) top surface plane) plays an important role in reducing or eliminating during the IV electrical characterization measurements, the potential barrier. The latter turns out in fact to be linked to the resulting resistivity of the electrodes as well. We also investigate the effect of thermal annealing of the diamond samples not only on the potential barrier but especially on the resistivity of the fabricated micro-wires. We shall see that the electrodes fabricated in a sample with (110) orientation do not exhibit any kind of barrier irrespective of the pulse energy and pulse duration used during the laser fabrication. Moreover, electrodes fabricated in the picosecond regime show no trace of barrier throughout a wide range of pulse energy values. Furthermore, the correlation between potential barrier and conductivity variation with respect to various laser writing parameters is also discussed in detail, and we report some of the lowest resistivity values that, compared to the literature, can be achieved with laser micromachining techniques.

# 2. Materials and Methods

## 2.1 Micromachining set-up and samples

The fabrication of conductive micro-electrodes perpendicular to the diamond surface was performed by means of a 20-Hz Ti:Sapphire amplified laser system (Amplitude) which delivers 40-fs transform-limited pulses at 790 nm wavelength in the mJ pulse energy range. Thanks to the laser compressor which can be detuned to change the pulse duration, it is possible to work at both femtosecond and picosecond regimes. This is crucial since pulse duration can influence the barrier height in the conductivity of an electrode as shown later in section 3.2. The results reported in this paper are related to a pulse duration of 200 fs and 10 ps respectively. As described in detail in [22], a 5 mm (full width at half maximum) Gaussian beam (with linear polarization) is converted into a Bessel beam (BB) after passing it through a fused silica axicon with a 178° apex angle. Thanks to a telescopic system formed by a 250 mm focal length lens ($L_1$) and a 0.45 N.A. 20× microscope objective ($f_{obj}$ = 9 mm), the final beam at the sample position is featured by 12° cone angle, a central core size of about 2.7 μm and a total Bessel zone of 700 μm. As shown in Fig.1, the Bessel beam, whose final non diffractive length is on the order of the diamond thickness, was sent orthogonally towards the sample placed on a micrometer precision 3D motorized stage (driven by SCA, system control application software, Altechna Rnd, Vilnius, Lithuania). The microfabrication of the graphitic wires was thus performed in a transverse configuration and without any sample translation.

Note that in the micromachining experiments the pulse energy was varied in a range such that the BB central core is the only portion of the beam which undergoes nonlinear absorption and interacts with the diamond sample, leading to microstructures with transverse size in the micrometer range. For a precise adjustment of the energy, a motorized attenuator consisting of a polarizer and rotating halfwave plate (Watt Pilot, Altechna Rnd, Vilnius, Lithuania) was placed just after an optical shutter used to select the number of laser pulses.

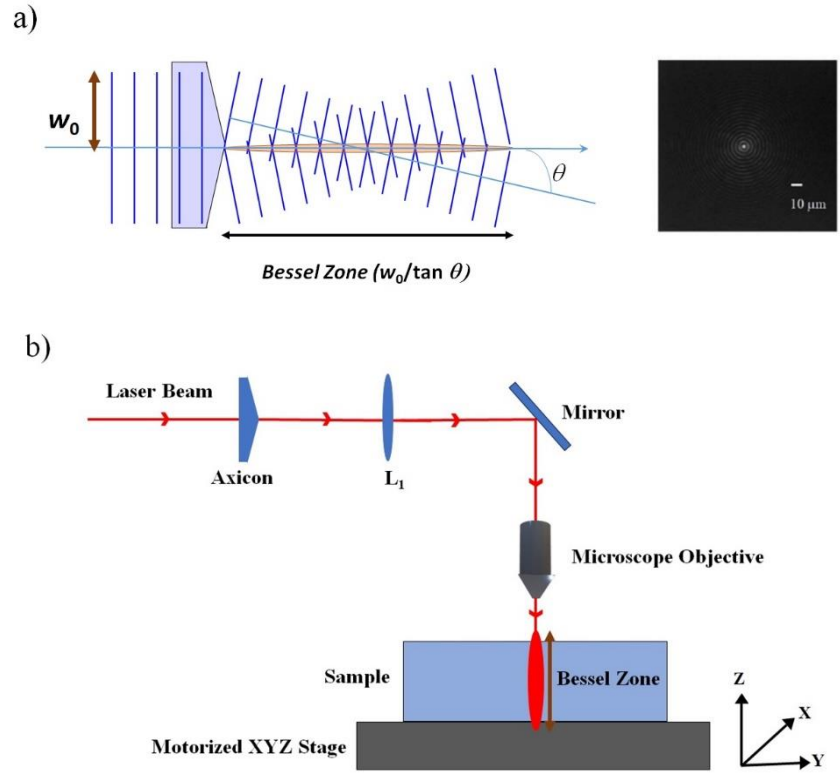

Fig:1 a) Schematic representation of the Bessel beam formation by an axicon with the transverse intensity profile recorded by the CCD at the sample position, b) Micromachining set up for a transverse Bessel beam writing of the sample.

As described in [22], a real-time observation of the sample surfaces (both top and bottom) was performed during the micromachining thanks to a backlighting Light Emitting Diode (LED) placed below the transparent sample and an imaging system using a lens and a CCD (not shown in the figure). This helps in setting up the relative positioning of the Bessel beam non-diffracting length with respect to the sample thickness, in order to have a symmetrical intensity distribution of the beam core across the sample to guarantee the formation of complete through-wires. As already discussed in [22], this also prevents strong and destructive interaction at the air/diamond interface where beam reshaping occurs because of the sudden change in the refractive index [24].

The diamond sample was placed on the motorized stage and aligned in both *x* and *y* axis (thanks to micrometer screws of the support) to make sure it is flat (within an error limit of 2 μm), in order to fabricate orthogonal wires with respect to the sample top and bottom surfaces. The Bessel beam micromachining was performed in multiple shot regime and, as already mentioned and shown in Fig. 1, the laser pulses were injected orthogonally to the sample surface. During the laser exposure the graphitic electrode starts growing from the bottom surface towards the top. It has been found that around 600 pulses are required to grow an electrode of length 500 µm in 10 ps regime and a minimum of 7000 in case of 200 fs regime. We believe this disparity arises from the fact that because of the higher intensity for a given energy, of a fs pulse with respect to the ps one, the damage on the top surface of the sample which is larger, can act as a barrier and lead to distortions for the incoming beam, in addition to the fact that the formed graphite at the air material interface will absorb part of the light. Nevertheless, due to its inherent conical energy reservoir for its formation, the Bessel beam gets recreated after a small distance depending on the size of the damage [25]. The fact that the BB undergoes an interruption could be the reason for which, in the fs regime, it takes more pulses for the graphitic wire to fully grow from bottom to the top of the sample. We mention that the growth rate of graphitic electrodes within the diamond bulk will be addressed in a future work. All the electrodes reported in this work have been created by sending 9000 Bessel pulses on the sample in both fs and ps laser machining regimes.

Here, we have explored the role of different laser writing parameters and of the crystallographic orientation on the potential barrier and thus on the conductivity of the graphitic wires fabricated. For this purpose, two samples were considered – a monocrystalline type IIa CVD diamond with (100) orientation and another sample of the same type with a (110) orientation, both produced by MB Optics. The dimensions of the samples are 5 mm × 5 mm × 0.5 mm.

## *2.2 Current-voltage measurements*

The electrical characterisation of the fabricated electrodes (featured by voltage-current measurements) was carried out by using conductive layers on the top and bottom surfaces of the diamond samples [22]. In particular, on one surface a silver metal deposition was carried out covering the end of the graphitic electrode (one end of the graphitic wires). In order to isolate one electrode from the others in order to avoid a short circuit during the electrical measurements, a mask was created by milling aluminium film (15 µm of thickness) with a nanosecond-pulsed Nd:YAG laser emitting in the IR (1064 nm), VIS (532 nm) or UV (355 nm) [26]. The mask dimensions were 250 × 250 µm$^2$ in area. Metal deposition for the fabrication of the electrodes was carried out using a thermal evaporator. The thickness of the deposition was 400 nm. The other side of the sample was fixed on a silicon wafer using silver paste thus completing the whole circuit (See [22] for details).

The current-voltage characterization was then conducted with a 2-probe configuration (tungsten-carbide microprobes controllable with micrometric accuracy) in a custom probe station. The probes were connected to a high-precision KeithleyTM 6487 voltage source, communicating with a computer through a LabVIEW system, which, in turn, measures the I-V curves. A "Leica Essential M50" optical microscope with an achromatic objective with magnification 0.63×, focal length of 148.2 mm and an eyepiece magnification of up to 100× interfaced with the I-V chamber allows for the accurate positioning of the microprobes on the sample.

*2.3 Micro-Raman spectroscopy*

Micro-Raman spectroscopy using a Labram Aramis John Yvon Horiba system with a DPSS laser source at 532 nm, was performed in order to characterize the effective transformation of diamond into amorphous or graphitic-like carbon (featured by the presence of $sp^2$ hybridized carbon) in the laser-written microstructures. The system was also equipped with a confocal microscope, an 1800 line mm$^{-1}$ grating, and an air-cooled CCD camera. The Raman signal collected in a backscattering configuration with a 100× objective, can achieve a spatial resolution of about 1 μm.

*2.4 Thermal annealing*

In the study presented here, the role of thermal annealing on the potential barrier change or the resistivity change was also investigated. To this end, the diamond samples were annealed thermally in ultra-high vacuum at 950º for 1 hour. The home-made annealing chamber consisted of a heating stage (HTR1002) made up of a unique combination of a dielectric ceramic material (Pyrolytic Boron Nitride) and an electrically conductive material (Pyrolytic Graphite). The dimensions are 50 mm × 60 mm with a thickness of around 1.40 mm [27]. It works in high vacuum condition reaching up to $5 \times 10^{-8}$ mbar; the heating process starts only at $5 \times 10^{-6}$ mbar. The heating stage is equipped with a type C thermocouple to read the actual temperature and a cover for heat shielding. The heater controller model is HC3500 by Tectra.

Note that the IV curves shown in this paper refer to the electrical characterization of the graphitic electrodes performed after annealing of the diamond samples.

# 3. Results and Discussions

In this section, we present the results of the Bessel beam micromachining of two samples of monocrystalline synthetic diamond featured by (100) orientation and (110) orientation respectively. In addition to study the effect of different laser writing parameters such as pulse energy and pulse duration, we shall focus on understanding the role of the crystallographic orientation on the morphology, the structural modification, and the IV characteristics (and on the presence or absence of a potential barrier) of the graphitic microwires obtained. We will thus compare the obtained results in the two different types of samples. In both cases, the electrode formation in the bulk follows the same mechanism. Under Bessel beam irradiation in the multiple shot regime, the formation of a graphite globule ($sp^2$ carbon) gets initiated at the bottom surface within the bulk. This can be explained by considering the reduced damage threshold and a higher local beam intensity at this position thanks to the constructive interference of the beam due to Fresnel reflections occurring on each diamond-air interface [22, 24]. Once the first transition of diamond to $sp^2$ carbon is completed, the process continues as a graphitization wave until it reaches the top surface since the incoming pulses get absorbed by the already transformed portion [12]. By injecting adequate number of pulses, electrodes of any length can be drawn within the diamond bulk.

## 3.1 General Morphology by optical characterization

The study of the electrodes fabricated in two differently oriented diamond crystals started with an analysis of the morphology of the 500 μm long graphitic wires obtained. The Bessel beam laser writing was performed in two different pulse duration regimes (femtosecond and picosecond regimes), and with the purpose of exploring the extremities of this chosen range, we used 200 fs and 10 ps pulses. The energy was varied in the micro Joule range. The structure and morphology of the microstructures obtained were observed under an optical microscope. In Fig. 2 in the top row, we show the wires fabricated with 200 fs pulses and a pulse energy of 2.5 μJ both in the (100) and (110) oriented crystals. In the bottom row we present the wires fabricated with 10 ps pulses and pulse energies of 2.5 μJ and 5 μJ, for the (110) oriented sample and the (100) oriented sample respectively. Note that the pulse energies chosen here are the threshold values for the generation of a complete graphitic microstructures in the two samples for the pulse durations used.

We can notice that the wires fabricated with 200 fs pulse duration (Figs. 2(a) and (b)), are featured by less cracks in comparison to those fabricated with 10 ps pulses (Fig. 2(c) and (d)), for both (100) and (110) oriented samples. The cracking effect at higher pulse durations can be attributed to the decrease in density of the transformed region compared to the untouched diamond. When a portion of the diamond lattice is converted to a graphitic-like phase under laser irradiation, its density becomes lower, resulting in a volume increase at that point and thus in an internal strain formation [28]. Since the diamond graphitization process takes just a few picoseconds [29], a graphitic globule of minimal dimension is formed without collateral damage in the case of wires fabricated with 200 fs short pulses. On the contrary, a large amount of energy is still deposited even after the formation of the first globule in case of higher pulse durations (such as the 10 ps pulses) leading to an enhanced cracking since the local strain overcomes the tensile strength of the diamond due to volume increase. It is worth noting that even when there is a cracking effect, the dimension of the main lobe remains the same irrespective of the pulse duration. In addition, the preferential cracking direction of the electrode towards (111) orientation, when machining a (100) cut diamond crystal, has already been observed and discussed in [30], and is attributed due to the lowest cleavage energy along that plane. In addition, as the energy increases, the length of the crack also increases irrespective of pulse duration [22].

If we focus on the differences in the morphology of the electrodes fabricated in the two differently oriented crystals, we can observe that the transverse size (≈ 1 μm) of the graphitic wires fabricated in the (110) oriented diamond crystal is smaller than that (≈ 2.5 μm) of those fabricated in a (100) oriented one, irrespective of the chosen energy or pulse duration. This can be noticed in the images of Fig. 2. In addition, the graphitic wires are more uniform with comparatively less cracks in the (110) cut sample as it is evident from Figs. 2(b) and 2(d).

The difference in the cracks direction along the wires fabricated in (100) and (110) cut samples has been explained by preferable cracking of the diamond crystal along (111) planes, which are turned at different angles depending on the chosen direction of observation [13]. When the diamond sample is irradiated orthogonally to the (100) crystallographic plane, the (111) planes are placed at an angle of 35.5º with respect to the direction of the laser beam (as confirmed by the cracks direction in Fig. 2(c). On the other hand, when the diamond sample is irradiated through the (110) crystallographic plane, the cleaved (111) planes are placed either parallel or perpendicular to the beam (see for instance later on the resulting cracks in Fig. 4).

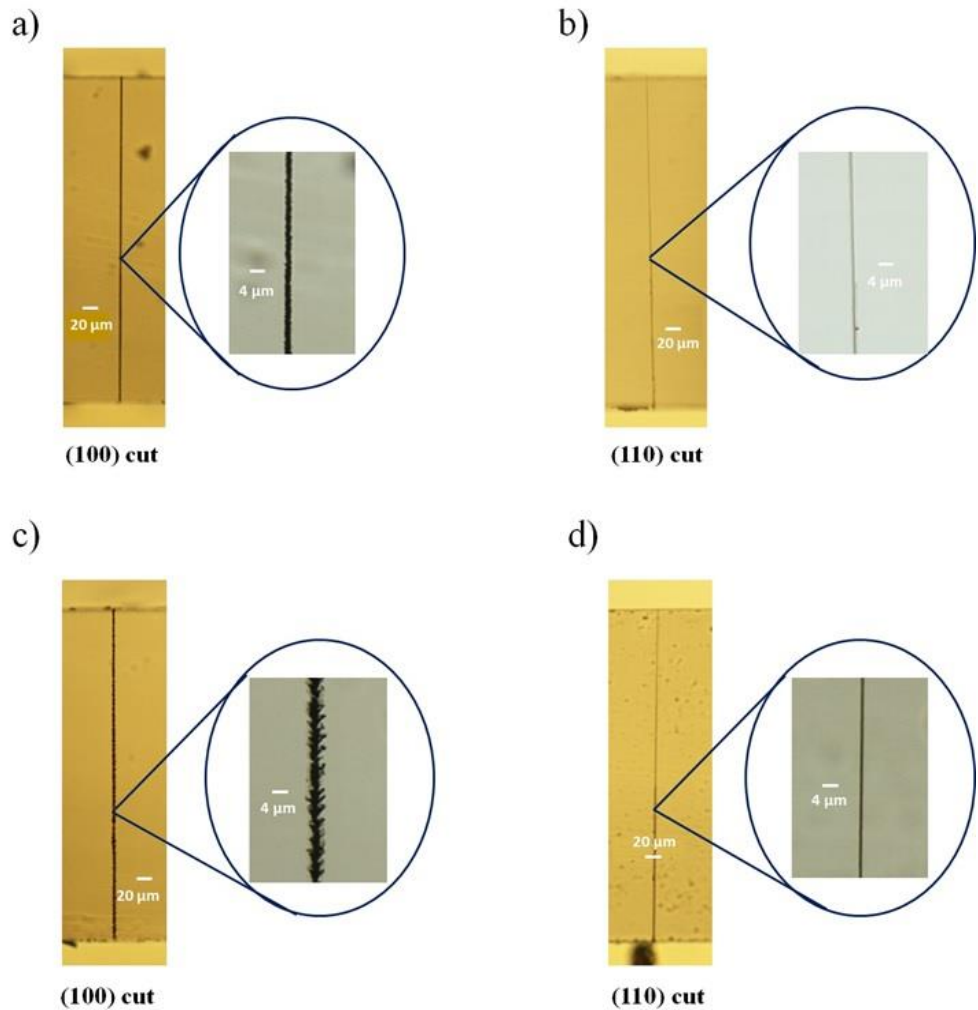

Fig. 2 Optical microscope images and relative zoom of the graphitic wires fabricated by Bessel beam with 200 fs (top row) and 10 ps pulses (bottom row) in a (100) cut diamond sample ((a) and (c) respectively), and in a (110) cut sample ((b) and (d) respectively). The pulse energy used here was the one just above minimum threshold value for the formation of the graphitic wires, namely 2.5 µJ for (a), (b) and (d), and 5 µJ for (c).

This results in smoother graphitic microstructures in the (110) oriented crystal especially at lower energy and lower pulse duration (Fig. 2.b). In this case, thanks to an energy close to the threshold, even though the cracking phenomena exists, the cracks are typically aligned in the direction of the electrode growth thus making it a single line without collateral damage. However, it was observed that at higher fabrication energies, the cracking effect is more pronounced. In fact, the laser writing experiments performed in this study showed that the morphology of the electrodes fabricated in the (100) and (110) oriented diamonds is different with respect to their continuity along the sample thickness. In particular, when increasing the pulse energy of the Bessel beam, we noticed that while in the first case ((100) orientation) the wires keep a continuous nature, the wires fabricated in the second case ((110) orientation) are featured by the presence of different segments separated by microcracks or microexplosions (leading to graphitic globules) along the whole microstructure. This phenomenon has been reported by Pimenov et.al. [31]. While the electrode formation is generally defined on the basis of graphitization wave [12], it also includes self-sustaining propagation of ionization wave as well. This means, while working in the multiple shot regime, the modification at a particular point in the diamond bulk can also lead to ionization of defects and the resultant seed electrons can be injected to the neighbouring undamaged region. These seed electrons can kick start an avalanche ionization and initiate transformation of that unmodified region even when the laser beam is absent. Therefore, the graphitized microstructure continues to grow into the bulk of diamond until the laser intensity at the front of the growing structure becomes less than a certain threshold value required for optical breakdown; therefore, at that point the ionization process together with the production of seed electrons stop. Figure 3 shows some of the wires fabricated in the (110) cut diamond sample both for 200 fs (Figs. 3(a) and (b)) and 10 ps pulse durations (Figs. 3(c) and (d)), and for two different pulse energy values higher than the material modification threshold value. It is worth mentioning that even though these electrodes seem to be featured by different segments (in contrast to those presented in Fig. 2 (b) and (d)), they turned out to be well connected (as confirmed by the IV characterisation we then performed and whose results will be shown later on).

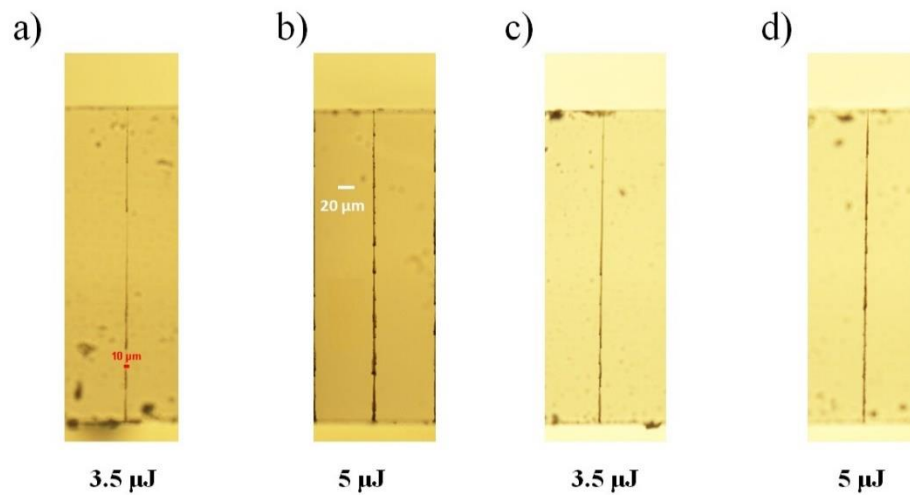

Fig. 3. Optical microscope images of the graphitic wires fabricated by Bessel beam in a (110) cut diamond sample with 200 fs pulse duration ((a) and (b)) and 10 ps pulse duration ((c) and (d)), showing the irregular nature of the microstructures along their length. The scale bar clearly shown in (b) is the same for all the presented images.

Finally, at even higher energies, the wires fabricated in the (110) cut sample turned out to be completely discontinuous. Instead of having a fully grown electrode, they formed a discrete set of graphitic modifications lines or isolated globules, as shown in Fig. 4 for the case of a microstructure obtained with 7 µJ and 200 fs Bessel pulses. Notice, that the presence of discrete globules in the fabrication was already observed by Kononenko et. al [13,32] in the case of standard laser writing with sample translation and explained in terms of a phenomenon called "incubation effect" occurring at higher machining fluences.

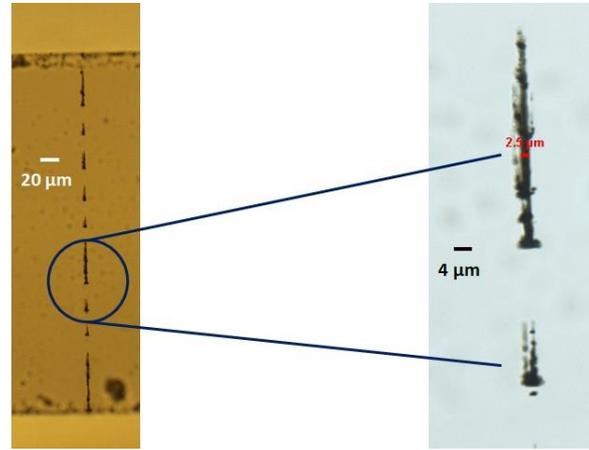

Fig. 4. Optical microscope image and relative zoom of a segmented microstructure fabricated in a (110) cut diamond sample by a 200 fs Bessel beam with pulse energy of 7 µJ.

On the other hand, nonlinear effects and self-focusing of the Bessel beam propagating in amorphous transparent materials have already been studied in the literature [33], but have not yet deeply investigated for crystals, although possible effects of a Gaussian beam self-focusing in diamond were hypothesized in [32]. In our case, we believe that what occurs in this high energy fabrication regime in a (110) cut diamond could be also the result of the high intensity distribution of the BB core in the central region of the sample thickness, favouring from local initial cracks (because of the orientation of the (111) plane), the separate growth of independent graphitic segments along the total Bessel focal zone. An investigation of the plasma distribution in diamond, formed during the radiation-matter interaction along the Bessel zone, and in different energy regimes, is the object of a separate and future experiment.

## 3.2 Electrical and structural characterization

A detailed study of the electrical properties of the fabricated in-bulk electrodes has been conducted, before and after thermal annealing of the diamond samples, with a focus on the evolution of the potential barrier height with respect to the different crystallographic orientations and beam parameters. Around twenty graphitic wires were fabricated on two CVD diamond samples with respectively (100) and (110) crystallographic orientation, by varying the pulse energy between 1 to 10 µJ, and with two different pulse durations (200 fs and 10 ps); and their resistance/resistivity was measured with a current–voltage measurement set-up as detailed in the methodology section. For all the tests mentioned in this work, we used a voltage range of -450 to 450 V, a compliance current of 25 mA, a step value of 5 V and a measurement time of 300 ms. The current-voltage (IV) tests for each electrode were conducted multiple times resulting in a standard deviation error of 5 %.

*Femtosecond fabrication regime*

In Figure 5, we show for illustration the measured IV curves measured after thermal annealing, related to two electrodes fabricated with a pulse energy of 3.5 µJ and a pulse duration of 200 fs, for the two different crystal orientations (100) and (110) respectively. We observe a clear difference between the curves of Fig. 5(a) and Fig. 5(b). The electrode fabricated in the first case ((100) orientation) shows the presence of a potential barrier close to 20 V which, in contrast, is absent in the second case ((110) orientation). In addition, the resistance of the former electrode is 420.5 kΩ while the latter has a resistance of 117 kΩ, which is significantly lower. In this case the resistivity turns out to be equal to 0.018 Ω cm, in contrast to the resistivity value of 0.41 Ω cm for the analogous electrode fabricated with same laser parameters in the (100) cut sample.

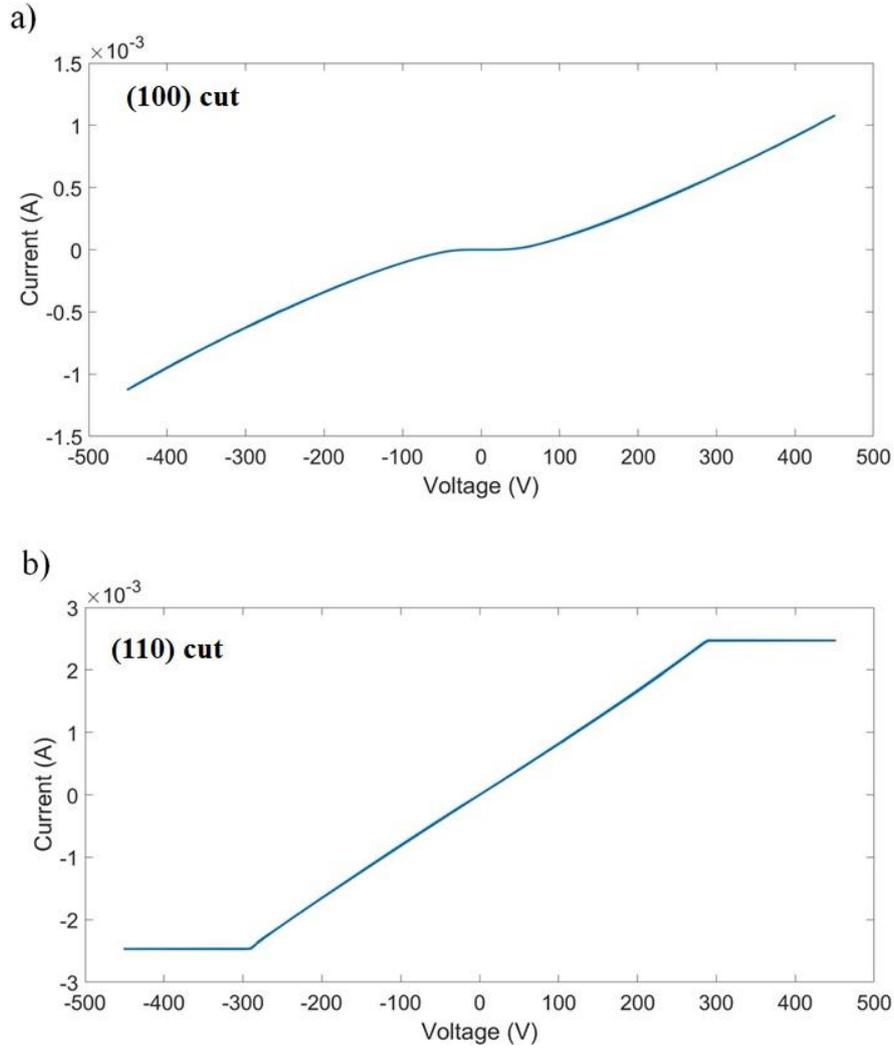

Fig. 5 Measured IV graphs after thermal annealing of graphitic electrodes respectively fabricated in a (100) oriented (a) and a (110) oriented (b) diamond crystal with a Bessel beam featured by a 200 fs pulse with energy of 3.5 µJ.

The improved conductance and the absence of a barrier for the electrode fabricated orthogonally to the (110) surface with respect to the one fabricated orthogonally to the (100) surface indicates, as a result of the BB machining with given pulse duration and pulse energy, that in that case a better transformation of diamond into $sp^2$ carbon occurs. We believe this is due to a higher heating of the diamond bulk in the (110) cut sample compared to (100) cut sample, leading a to better breakage of $sp^3$ into $sp^2$ phase. Generally, the simplest and general explanation for the intensive diamond heating which results in the transformation and creation of $sp^2$ carbon is the dissipation of energy absorbed at the front of the modified region progressively growing from the bottom to the top surface. The key factor affecting the diamond graphitization under laser irradiation is diamond cracking in front of the modified region. The cracks provide structural defects activating the graphitization process within the heated diamond region. In case of (110) oriented diamond crystal, an additional heating mechanism has been reported by Kononenko et al. [13]. This, also known as enhancement of direct plasma-assisted absorption of laser radiation in diamond due to intensified local diamond ionization, has a positive impact on diamond ionization as a result of local electric field enhancement in the microcracks generated in the front of the graphitization wave in addition to the normal heating. The reason for this secondary heating effect lies in the fact that when the graphitic wire is fabricated orthogonally to the (110) plane, the crack tip is situated close to the beam axis, where the Bessel core intensity can reach the maximum value. The appearance of long cracks in the direction of beam axis may have a resonant effect thus increasing the diamond ionization. As a consequence, this should result in increased graphitization rate and thus, better transformation of diamond into amorphous carbon and graphitic phases.

Moreover, it is worth noting that even if the laser modified portion of the diamond bulk looks uniform and continuous (as the microstructures shown in Fig.2), there is a higher chance that in the microscopic regime, that may not be the case. Instead, there could be some microscopic gaps between each graphitic globule and thus, to have a continuous path for the charge transfer, these gaps should be overcome [23]. The voltage required to break down the gap is the potential barrier/breakdown voltage, which we observe in Fig. 5(a), where the IV curve for the electrode fabricated in the (100) cut sample is reported. In contrast, in the case of the electrode fabricated in the (110) cut sample, the microscopic gaps between the graphitic globules are reduced to a great extend thanks to the greater heating mechanism discussed above. Also, as said above, the fact that the diamond cracking during the graphitization process occurs in the beam propagation direction, i.e. along the wire, helps in enhancing the structural transformation of diamond into $sp^2$ carbon, as confirmed by the higher conductivity of the electrode. The same behaviour has been observed for a wide range of pulse energies used in the fabrication of wires orthogonal to the (110) face, as no trace of barrier potential has been observed in any of the electrodes fabricated with our laser parameters (as reported also later on in table 3).

In order to investigate the crystalline structural changes and the extend of the material transformation after the laser-matter interaction, micro-Raman spectroscopy was performed on the wires fabricated in the two diamonds featured by different crystallographic orientation. The laser beam used for the measurement was focused at the top of the electrodes in correspondence with the centre of the transverse section of the wires extremity, where we expect a higher concentration of graphite, as previously discussed [22]. In Figure 6, we present the spectra recorded in the central part of two wire cross-sections corresponding to two electrodes fabricated with a pulse energy of 3.5 µJ at 200 fs in the (100) cut sample (a) and in the (110) cut sample (b) respectively. In both cases it is possible to observe three characteristic peaks: a sharp, narrow peak at 1332.2 cm$^{-1}$ (corresponding to diamond), and two smaller, broader peaks centered at ~ 1335 cm$^{-1}$ and ~1590 cm$^{-1}$ corresponding respectively to graphite D ("disorganised") and to graphite G ("organised") respectively [34, 35]. In order to better understand the effect of the laser micromachining on the two different diamond samples, the Raman spectra have been "deconvolved" considering the contribution of diamond and of amorphous carbon phases through a fitting procedure as described by A. Dychalska et al. [36].

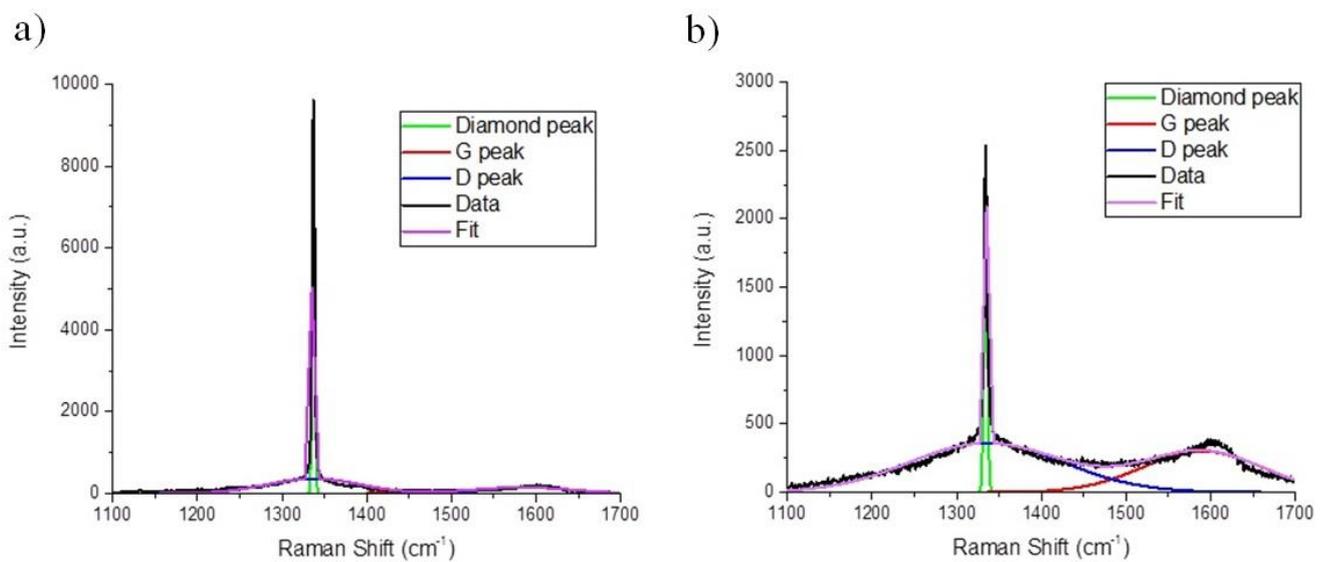

Fig. 6 Micro-Raman spectra recorded at the top of two electrodes fabricated with 200 fs pulse duration and 3.5 µJ energy in a (100) cut sample (a) and a (110) cut sample (b) respectively.

The presence of additional peaks (D (1335 cm$^{-1}$)) and G (1590 cm$^{-1}$)) together with the diamond peak (1332.2 cm$^{-1}$) in case of both electrodes fabricated suggest that there is a conversion from sp$^3$ to sp$^2$ carbon phases thus transforming the diamond into a mixture of amorphous carbon and graphite through laser writing irrespective of the sample orientation. On the other hand, the two spectra reported in the figure present some differences. In the case of the wire created orthogonally to the (100) plane, the intensity of the diamond peak remains high while it is suppressed in the case of the wire fabricated orthogonally to the (110) plane. The opposite happens for the G-peak which correlates to the presence of graphite. The G-peak is slightly more pronounced for the latter electrode compared to the former one. From the spectra, it is possible to estimate the ratio of the graphite G peak to the diamond peak which, in turn, can give a measure of the graphite content inside the electrodes fabricated in two samples with different orientations [37]. This ratio is higher for the electrode of the (110) cut sample, while it is lower for the electrode of the (100) cut sample. Therefore, the indication of a higher content of graphite in the (110) oriented crystal is consistent with the greater electrode conductivity (and possibly with the absence of potential barrier) highlighted before from the IV measurements. Also note that the intensity ratio between the disordered D peak and the graphite G peak, I(D)/I(G), being less than one for the spectrum relative to the electrode fabricated orthogonally to the (110) sample, indicates that in that case there is more graphite portion than amorphous carbon in its $sp^2$ state [3].

*Picosecond fabrication regime*

The electrical characterization of the graphitic wires fabricated with 10 ps pulse duration was performed for a comparison with the results presented so far. None of the resulting electrodes showed a trace of potential barrier in the resistivity measurements, at any pulse energy regime, and in both crystallographic orientations (in accordance with the results of [22] presented in the (100) cut sample case). In Figure 7, we present IV curves related to wires fabricated with 10 ps pulse duration and with two different pulse energies in both (100) and (110) oriented samples. The lower energy values chosen for each crystallographic orientation correspond to the minimum energy thresholds for the graphitic wire formation (as also mentioned in relation to Fig. 2).

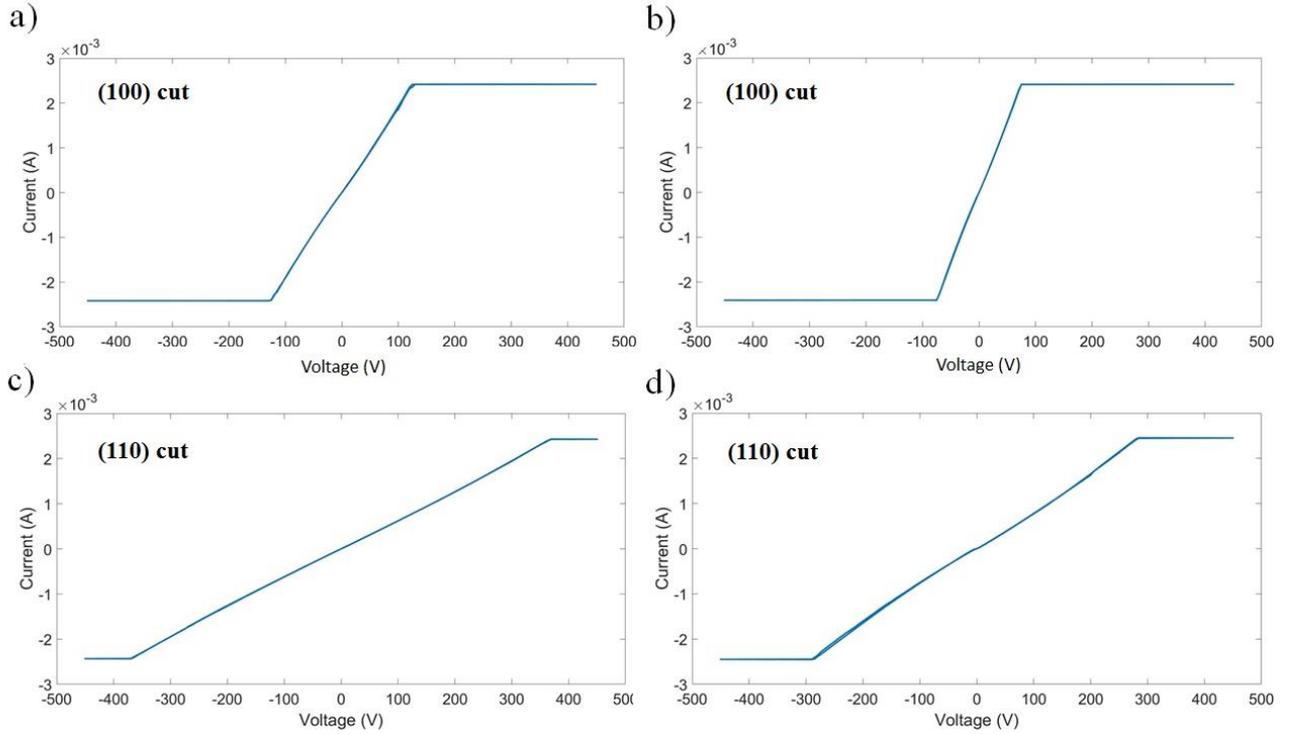

Fig. 7 Measured IV graphs of electrodes fabricated at 10 ps with a pulse energy of 5 µJ (a) and 10 µJ (b) in the (100) cut sample, and with a pulse energy of 2.5 µJ (c) and 3.5 µJ (d) in the (110) cut sample.

The resistances related to Figs. 7 (a) and (b) are found to be 50 kΩ and 31 kΩ while those related to (c) and (d) are 151 kΩ and 116 kΩ respectively. It is worth noting that the pulse energy used for the generation of the wires orthogonal to the (100) plane is higher than in the other case, and this explains the lower resistance values shown by the IV curves of the electrodes fabricated in the (100) oriented diamond sample, as we have observed that the conductance increases when increasing the Bessel beam pulse energy (up to a certain value) [22]. Therefore, note that a direct comparison of resistance and resistivity values between wires fabricated in this picosecond regime in the two differently oriented samples, is not possible for fixed laser writing parameters. It is also worth mentioning that it is more challenging to fabricate well connected electrodes with higher pulse energies in (110) cut samples, as the graphitic microstructures formed in the diamond bulk in that case are then featured by segmentation, as discussed in section 3.1. The absence of any potential barrier for any pulse energy regime at 10 ps, for both crystal orientations, may be explained by recalling that a better transformation of diamond into $sp^2$ carbon occurs in case of longer pulse durations [22, 38].

## 3.3 Role of pulse energy

As it has already been established that the pulse energy used in the micromachining process can influence the conductivity of the electrode obtained [22], in this work, we explored in particular the role of the Bessel pulse energy in the appearance of a barrier potential, comparing the results in the two differently oriented diamond samples ((100) cut and (110) cut). For this study we concentrated our attention on graphitic microstructures fabricated in the femtosecond pulse regime (200 fs pulse duration) where the presence of a barrier potential at least in the (100) cut sample was already observed (as shown before). Pulse energy values ranging from 1 µJ to 6 µJ were chosen for the fabrication of the graphitic wires in both samples (note that above 6 uJ in the (110) cut sample we do not have continuous wires). In Figure 8, we report the evolution of the potential barrier height with respect to the pulse energy used, retrieved from the IV measurements performed.

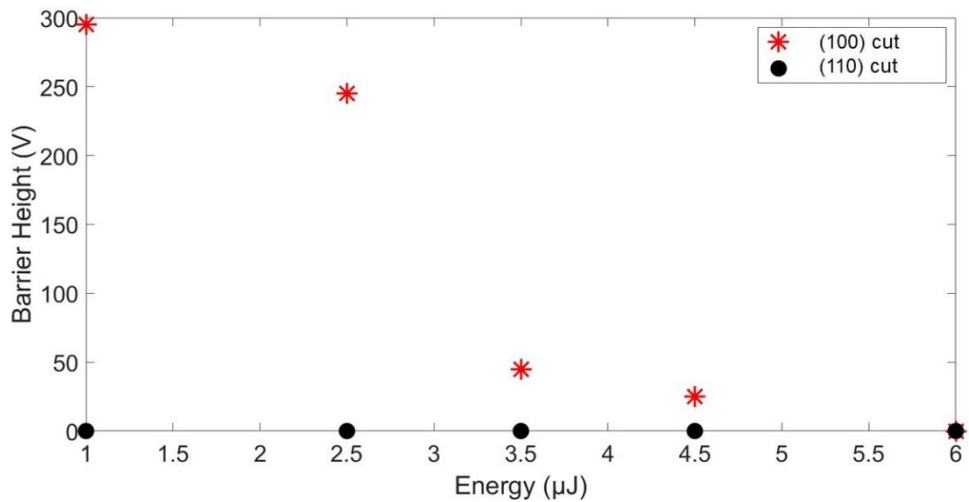

Fig. 8 Evolution of the potential barrier height with respect to the pulse energy used in the fabrication of the microelectrodes with 200 fs Bessel pulses in two samples with (100) and (110) crystallographic orientation respectively. The measurements error bars are within the size of the symbols used.

It is well evident that the electrodes fabricated respectively in the (100) and (110) oriented samples behave differently when it comes to the dependence of the barrier height with respect to the pulse energy used for the laser fabrication. While the low energy electrodes fabricated orthogonally to the (100) plane show a huge potential barrier in the electrical characterization, this is nil for those fabricated orthogonally to the (110) plane. In fact, in all cases, for the wide range of energy values investigated in our experiment, the electrodes fabricated in this second sample showed no trace of barrier potential. At the same time, for the electrodes created in the (100) cut sample, we observe a trend in the evolution of the potential barrier height, the latter being very high at low pulse energies but then progressively decreasing for increasing energy values, till disappearing completely at 6 µJ.

A similar trend is observed in the evolution of the resistivity values that can be extracted from the IV measurements. Table 1 summarizes the resistivity evaluated for the graphitic microstructures fabricated at different pulse energies in both diamond samples orientations.

| Energy (µJ) | Resistivity (Ω cm) – (100) cut | Resistivity (Ω cm) – (110) cut |
|---|---|---|
| 1 | 169 | 0.044 |
| 2.5 | 6.3 | 0.043 |
| 3.5 | 0.41 | 0.018 |
| 4.5 | 0.3 | 0.018 |
| 6 | 0.128 | 0.013 |

Table 1: Evolution of resistivity with respect to pulse energy of microelectrodes fabricated with 200 fs Bessel pulses.

The resistivity of the electrodes fabricated in the (100) oriented crystal follows the same trend as the barrier potential height itself, decreasing for increasing pulse energies. Notice that for 1 µJ both the barrier potential height and the resistivity are featured by extremely high values. On the other hand, in the case of the (110) oriented sample, the resistivity values are low irrespective of the energy values used for the laser fabrication, also in line with the fact that in this case no potential barrier was revealed. A resistivity decrease as the energy is increased, is also observed.

Although we know that the femtosecond regime of laser machining in diamond generally leads to graphitic wires with lower conductivity compared to long pulse durations [22], here for a (110) cut diamond sample we proved that in this regime it is possible to obtain low resistivity values, in particular a resistivity of 0.013 Ω cm, as in the case of the microelectrode fabricated with a Bessel pulse of 200 fs duration and 6 µJ energy. Finally note that the lower resistivity values measured in the (110) oriented sample are not only due to the smaller core size of the graphitic wires fabricated in the latter, but also to the resistance values retrieved from the IV measurements, which turned out to be systematically smaller than those relative to the wires of the (100) oriented sample, for a given pulse energy. As an example, from the IV curves reported in Fig. 5 and corresponding to microelectrodes fabricated with 200 fs pulses and energy of 3.5 µJ, we find a resistance value of 420 kΩ in the (100) substrate, and 117 kΩ in the (110) substrate, respectively."

## 3.4 Role of thermal annealing

As also stated at the end of section 2, the results just presented refer to the characterization of the fabricated electrodes in the diamond samples, after thermal annealing of the latter.

In this section, we present for completeness a comparison of the potential barrier height and the resistivity values obtained before and after the thermal annealing of the samples. The results of the electrical characterization are summarized in Table 2 and Table 3, referring respectively to the electrodes fabricated in the (100) and (110) oriented diamond samples.

| Energy (µJ) | Barrier (V) before annealing | Barrier (V) after annealing | Resistivity ($\Omega$ cm) before annealing | Resistivity ($\Omega$ cm) after annealing |
|---|---|---|---|---|
| 1 | 295 | 295 | 200 | 169 |
| 2.5 | 245 | 245 | 14.7 | 6.3 |
| 3.5 | 45 | 45 | 0.44 | 0.41 |
| 4.5 | 25 | 25 | 0.31 | 0.3 |
| 6 | 0 | 0 | 0.135 | 0.128 |

Table 2: Evolution of potential barrier height and resistivity values of the electrodes fabricated by Bessel beam in a (100) cut sample at 200 fs for different pulse energies and characterized before and after thermal annealing.

| Energy (µJ) | Barrier (V) before annealing | Barrier (V) after annealing | Resistivity ($\Omega$ cm) before annealing | Resistivity ($\Omega$ cm) after annealing |
|---|---|---|---|---|
| 1 | 0 | 0 | 0.047 | 0.044 |
| 2.5 | 0 | 0 | 0.048 | 0.043 |
| 3.5 | 0 | 0 | 0.028 | 0.018 |
| 4.5 | 0 | 0 | 0.023 | 0.018 |
| 6 | 0 | 0 | 0.023 | 0.013 |

Table 3: Evolution of potential barrier height and resistivity values of the electrodes fabricated by Bessel beam in a (110) cut sample at 200 fs for different pulse energies and characterized before and after thermal annealing.

The results clearly highlight that the barrier potential height, when it exists, remains the same before and after annealing, irrespective of the pulse energy chosen or the sample orientation. The only variable that shows an evolution is the resistivity which tends to decrease after the thermal annealing of the samples. We notice though, that even in this scenario, the improvement of the conductivity is not uniform. The electrodes fabricated at low pulse energy which had really high resistivity values before annealing show a huge difference and present a lower resistivity after annealing, especially those for the (100) oriented sample. But, as the pulse energy used for the fabrication increases and the resistivity decreases, the variation after annealing gets reduced and eventually it becomes almost negligible. For the electrodes fabricated in the (110) cut sample, the change in the resistivity of the electrodes is not so pronounced. One of the possible explanations for this kind of trend could be that at low energy, as for the micromachining of the (100) cut sample, the transformation of the crystalline structure of diamond into $sp^2$ phase is only partial, and thus with the annealing it gets enhanced thanks to the thermal effects. But this process of better transformation may occur only in the regions where the partial graphitization already occurred during laser machining, before annealing. On the other hand, as the barrier potential is the voltage required to breakdown the gaps between nanocrystalline graphite sheets, this remains constant as the thermal annealing affects only the portion that has already been transformed. At the same time, as we use higher pulse energies in the laser writing, the graphitization process (with maximum conversion of $sp^3$ phase into $sp^2$ carbon) is already efficient, as already discussed, and therefore there may be only little or no room for annealing to have a visible effect on the conductivity, leading thus to a minimal resistivity difference (before and after annealing) in the case of electrodes fabricated at higher energies. As for the electrodes fabricated in the (110) cut sample, they behave as the electrodes fabricated in the higher pulse energy regime in the (100) cut sample, where an efficient graphitization process already occurs during the laser-matter interaction, leading to very small resistivity values of the wires and thus, negligible resistivity difference before and after annealing.

# Conclusion

In this work, the graphitic microelectrodes have been fabricated in monocrystalline 500 μm thick CVD diamond samples with orientations (100) and (110) respectively, by using a pulsed Bessel beam in the femtosecond and picosecond regime, impinging orthogonally to the sample surface. The optimization of the morphology and conductivity of the electrodes has been done by using different laser writing parameters. Different characterisation techniques such as optical microscopy, current-voltage tests and micro-Raman measurements have been used to analyse the morphological, electrical, and structural features of the electrodes.

We have shown that in addition to the use of specific pulse energies and durations, the crystallographic orientation of the sample can have a prominent influence in reducing or eliminating the potential barrier height of the laser written graphitic electrodes. With this study, the impact of the crystallographic orientation of the diamond samples on the overall quality of an electrode has been clearly demonstrated as the graphitic wires fabricated orthogonally to the (110) plane showed no trace of potential barrier in any energy regime or pulse duration, in contrast to those fabricated orthogonally to the (100) oriented-crystal case where the barrier is generally observed, especially when obtained with low pulse energy and 200 fs pulses. We believe that thanks to the additional heating mechanism that arises from the directionality of cracking during the fabrication of the wire, the possible microgaps that may exist between the graphitic globules that are generated along the beam path, are drastically reduced in (110) cut samples resulting, therefore, in this case, in high quality electrodes with good conductivity and a barrier potential height definitely negligible.

A comparison between the micro-Raman spectra recorded for wires fabricated in the fs regime in the two differently oriented samples indicated that there is a better transformation of diamond into graphite when the wires are fabricated orthogonally to the (110 plane). From the morphological point of view, we have shown that in addition of being thinner (1 μm size approximately), the electrodes fabricated in the (110) cut sample present a smoother core than those obtained in the (100) cut sample, especially at low energy regime. On the other hand, in the (110) oriented sample, micromachining with higher energies (larger than 6 μJ with the geometry of our Bessel beam) leads to the breakage of the graphitic microstructures, possibly due to incubation effects and also nonlinear propagation effects of the Bessel beam inside the bulk. This acts as a limit to explore a wide high energy regime for the generation of continuous electrodes throughout the whole sample thickness.

The effect of thermal annealing of the diamond samples, on the resistivity of the fabricated micro-electrodes has also been investigated. We found that while there is indeed a reduction in the resistivity after the annealing process especially for the wires fabricated with low energy pulses in both the samples, the barrier potential height which was present in the IV curves in the (100) oriented sample, remains, on the other hand, the same.

In the (110) oriented diamond, resistivities lower than 0.015 Ω cm have been obtained, namely 0.013 Ω cm for a 1 μm transverse size electrode, which, to the best of our knowledge, is one of the lowest values compared to the literature results and the lowest using Bessel beams. In addition, it is the lowest value achieved for an in-bulk graphitic micro-electrode written perpendicular to the surface of the sample by laser micromachining.

With this study we have found some particular conditions for an improvement of the overall quality of the laser written electrodes in terms of morphology (uniformity of the wires) and conductivity values. This opens the way to the fast fabrication by pulsed Bessel beams without any sample translation, of graphitic electrodes with different geometries and configurations that may find their applications not only for sensing in photonic and microfluidic chips, but also in the case of 3D diamond detectors for highly energetic radiation, to be used for instance in nuclear physics and medical dosimetry.

## Authorship contribution statement

O. J. conceived, supervised the micromachining experiments, discussed the results and revised the entire manuscript. A. K. realized the micromachining of the diamond samples, analysed the results, and wrote the first manuscript draft. Raman measurements were performed and analysed by A. C. The electrical measurements were performed by A. K under the supervision of P. A. All authors revised the manuscript.

## Declaration of competing interests

The authors declare no conflict of interest.

## Acknowledgements

O. J. would like to thank David Moran from Glasgow University, and all the authors would like to thank Federico Picollo, Adam Britel, and Paolo Olivero from Torino University for useful discussions, as well as Vanna Pugliese for helping with the annealing processes.

This research has received funding from the European Union's H2020 Marie Curie ITN project LasIonDef (GA n.956387) and QuantDia (FISR2019-05178) funded by Ministero dell'Istruzione, dell'Università e della Ricerca, and from the bilateral scientific cooperation project CNR-RS (UK) (2022-2023) entitled "Laser-patterned diamond superconducting single-photon sensors".